\RequirePackage{lineno}
\documentclass[aps,prl,twocolumn,superscriptaddress,showpacs,preprintnumbers,amsmath,amssymb]{revtex4}

\usepackage{afterpage,rotating}
\usepackage{currvita}

\usepackage{tabularx}
\usepackage{nonfloat}
\usepackage{graphicx}
\usepackage{dcolumn}
\usepackage{bm}
\usepackage{subfigure}
\usepackage{mathrsfs}
\usepackage{amsmath,amsbsy,amssymb,lmodern}
\usepackage{epsfig,color}
\usepackage{placeins}
\usepackage{rccol}
\usepackage{comment}
\usepackage{hyperref} 
\usepackage{setspace} 
\usepackage{tabularx} 
\usepackage{multirow}
\usepackage{braket}
\usepackage{accents}

\newcommand{\bseep}{B_s^0 \rightarrow \eta^\prime \eta}
\newcommand{\bksep}{B^0 \rightarrow \eta^\prime K_S^0}
\DeclareRobustCommand\mybar[1]{\accentset{\rule{0.6em}{0.6pt}}{#1}}
\newcommand{\mbc}{M_{\rm bc} }
\newcommand{\de}{\Delta E}
\newcommand{\mep}{M(\pi^+\pi^-\eta)}
\newcommand{\beep}{B_s^0 \rightarrow \eta^\prime \eta} 
\newcommand{\et}{\eta}
\newcommand{\bst}{B_s^{*0}}
\newcommand{\bs}{B_s^0}
\newcommand{\fbssbssall}{f_{B_{s}^{(*)0} \mybar{B}_s^{(*)0}}}
\newcommand{\fbssbss}{f_{B_{s}^{*0} \mybar{B}_s^{*0}}}
\newcommand{\fbsbss}{f_{B_{s}^{*0} \mybar{B}_s^0}}
\newcommand{\etp}{\eta^{\prime}}

\newcommand{\bsbs}{B_s^{*0} \mybar{B}_s^{*0}$, $B_s^{*0} \mybar{B}_s^0$, and $ B_s^0 \mybar{B}_s^0}

\newcommand{\bsbsbara}{B_s^{(*)0}\mybar{B}_s^{(*)0}}

\begin{document} 
%

%



\title{ \quad\\[1.0cm] Search for the Decay \boldmath$\bseep$}

\noaffiliation
\affiliation{Department of Physics, University of the Basque Country UPV/EHU, 48080 Bilbao}
\affiliation{University of Bonn, 53115 Bonn}
\affiliation{Brookhaven National Laboratory, Upton, New York 11973}
\affiliation{Budker Institute of Nuclear Physics SB RAS, Novosibirsk 630090}
\affiliation{Faculty of Mathematics and Physics, Charles University, 121 16 Prague}
\affiliation{Chonnam National University, Gwangju 61186}
\affiliation{University of Cincinnati, Cincinnati, Ohio 45221}
\affiliation{Deutsches Elektronen--Synchrotron, 22607 Hamburg}
\affiliation{Duke University, Durham, North Carolina 27708}
\affiliation{Department of Physics, Fu Jen Catholic University, Taipei 24205}
\affiliation{Key Laboratory of Nuclear Physics and Ion-beam Application (MOE) and Institute of Modern Physics, Fudan University, Shanghai 200443}
\affiliation{Justus-Liebig-Universit\"at Gie\ss{}en, 35392 Gie\ss{}en}
\affiliation{Gifu University, Gifu 501-1193}
\affiliation{II. Physikalisches Institut, Georg-August-Universit\"at G\"ottingen, 37073 G\"ottingen}
\affiliation{SOKENDAI (The Graduate University for Advanced Studies), Hayama 240-0193}
\affiliation{Gyeongsang National University, Jinju 52828}
\affiliation{Department of Physics and Institute of Natural Sciences, Hanyang University, Seoul 04763}
\affiliation{University of Hawaii, Honolulu, Hawaii 96822}
\affiliation{High Energy Accelerator Research Organization (KEK), Tsukuba 305-0801}
\affiliation{J-PARC Branch, KEK Theory Center, High Energy Accelerator Research Organization (KEK), Tsukuba 305-0801}
\affiliation{Higher School of Economics (HSE), Moscow 101000}
\affiliation{Forschungszentrum J\"{u}lich, 52425 J\"{u}lich}
\affiliation{IKERBASQUE, Basque Foundation for Science, 48013 Bilbao}
\affiliation{Indian Institute of Science Education and Research Mohali, SAS Nagar, 140306}
\affiliation{Indian Institute of Technology Bhubaneswar, Satya Nagar 751007}
\affiliation{Indian Institute of Technology Guwahati, Assam 781039}
\affiliation{Indian Institute of Technology Hyderabad, Telangana 502285}
\affiliation{Indian Institute of Technology Madras, Chennai 600036}
\affiliation{Indiana University, Bloomington, Indiana 47408}
\affiliation{Institute of High Energy Physics, Chinese Academy of Sciences, Beijing 100049}
\affiliation{Institute of High Energy Physics, Vienna 1050}
\affiliation{Institute for High Energy Physics, Protvino 142281}
\affiliation{INFN - Sezione di Napoli, 80126 Napoli}
\affiliation{INFN - Sezione di Torino, 10125 Torino}
\affiliation{Advanced Science Research Center, Japan Atomic Energy Agency, Naka 319-1195}
\affiliation{J. Stefan Institute, 1000 Ljubljana}
\affiliation{Institut f\"ur Experimentelle Teilchenphysik, Karlsruher Institut f\"ur Technologie, 76131 Karlsruhe}
\affiliation{Kavli Institute for the Physics and Mathematics of the Universe (WPI), University of Tokyo, Kashiwa 277-8583}
\affiliation{Department of Physics, Faculty of Science, King Abdulaziz University, Jeddah 21589}
\affiliation{Kitasato University, Sagamihara 252-0373}
\affiliation{Korea Institute of Science and Technology Information, Daejeon 34141}
\affiliation{Korea University, Seoul 02841}
\affiliation{Kyungpook National University, Daegu 41566}
\affiliation{Universit\'{e} Paris-Saclay, CNRS/IN2P3, IJCLab, 91405 Orsay}
\affiliation{P.N. Lebedev Physical Institute of the Russian Academy of Sciences, Moscow 119991}
\affiliation{Faculty of Mathematics and Physics, University of Ljubljana, 1000 Ljubljana}
\affiliation{Ludwig Maximilians University, 80539 Munich}
\affiliation{Luther College, Decorah, Iowa 52101}
\affiliation{Malaviya National Institute of Technology Jaipur, Jaipur 302017}
\affiliation{University of Maribor, 2000 Maribor}
\affiliation{Max-Planck-Institut f\"ur Physik, 80805 M\"unchen}
\affiliation{School of Physics, University of Melbourne, Victoria 3010}
\affiliation{University of Mississippi, University, Mississippi 38677}
\affiliation{University of Miyazaki, Miyazaki 889-2192}
\affiliation{Moscow Physical Engineering Institute, Moscow 115409}
\affiliation{Graduate School of Science, Nagoya University, Nagoya 464-8602}
\affiliation{Kobayashi-Maskawa Institute, Nagoya University, Nagoya 464-8602}
\affiliation{Universit\`{a} di Napoli Federico II, 80126 Napoli}
\affiliation{Nara Women's University, Nara 630-8506}
\affiliation{National Central University, Chung-li 32054}
\affiliation{National United University, Miao Li 36003}
\affiliation{Department of Physics, National Taiwan University, Taipei 10617}
\affiliation{H. Niewodniczanski Institute of Nuclear Physics, Krakow 31-342}
\affiliation{Nippon Dental University, Niigata 951-8580}
\affiliation{Niigata University, Niigata 950-2181}
\affiliation{University of Nova Gorica, 5000 Nova Gorica}
\affiliation{Novosibirsk State University, Novosibirsk 630090}
\affiliation{Osaka City University, Osaka 558-8585}
\affiliation{Pacific Northwest National Laboratory, Richland, Washington 99352}
\affiliation{Panjab University, Chandigarh 160014}
\affiliation{Peking University, Beijing 100871}
\affiliation{University of Pittsburgh, Pittsburgh, Pennsylvania 15260}
\affiliation{Punjab Agricultural University, Ludhiana 141004}
\affiliation{Research Center for Nuclear Physics, Osaka University, Osaka 567-0047}
\affiliation{Meson Science Laboratory, Cluster for Pioneering Research, RIKEN, Saitama 351-0198}
\affiliation{Department of Modern Physics and State Key Laboratory of Particle Detection and Electronics, University of Science and Technology of China, Hefei 230026}
\affiliation{Seoul National University, Seoul 08826}
\affiliation{Showa Pharmaceutical University, Tokyo 194-8543}
\affiliation{Soochow University, Suzhou 215006}
\affiliation{Soongsil University, Seoul 06978}
\affiliation{Sungkyunkwan University, Suwon 16419}
\affiliation{School of Physics, University of Sydney, New South Wales 2006}
\affiliation{Department of Physics, Faculty of Science, University of Tabuk, Tabuk 71451}
\affiliation{Tata Institute of Fundamental Research, Mumbai 400005}
\affiliation{Department of Physics, Technische Universit\"at M\"unchen, 85748 Garching}
\affiliation{School of Physics and Astronomy, Tel Aviv University, Tel Aviv 69978}
\affiliation{Toho University, Funabashi 274-8510}
\affiliation{Department of Physics, Tohoku University, Sendai 980-8578}
\affiliation{Earthquake Research Institute, University of Tokyo, Tokyo 113-0032}
\affiliation{Department of Physics, University of Tokyo, Tokyo 113-0033}
\affiliation{Tokyo Institute of Technology, Tokyo 152-8550}
\affiliation{Tokyo Metropolitan University, Tokyo 192-0397}
\affiliation{Utkal University, Bhubaneswar 751004}
\affiliation{Virginia Polytechnic Institute and State University, Blacksburg, Virginia 24061}
\affiliation{Wayne State University, Detroit, Michigan 48202}
\affiliation{Yamagata University, Yamagata 990-8560}
\affiliation{Yonsei University, Seoul 03722}
  \author{N.~K.~Nisar}\affiliation{Brookhaven National Laboratory, Upton, New York 11973} 
  \author{V.~Savinov}\affiliation{University of Pittsburgh, Pittsburgh, Pennsylvania 15260} 
  \author{I.~Adachi}\affiliation{High Energy Accelerator Research Organization (KEK), Tsukuba 305-0801}\affiliation{SOKENDAI (The Graduate University for Advanced Studies), Hayama 240-0193} 
  \author{H.~Aihara}\affiliation{Department of Physics, University of Tokyo, Tokyo 113-0033} 
  \author{S.~Al~Said}\affiliation{Department of Physics, Faculty of Science, University of Tabuk, Tabuk 71451}\affiliation{Department of Physics, Faculty of Science, King Abdulaziz University, Jeddah 21589} 
  \author{D.~M.~Asner}\affiliation{Brookhaven National Laboratory, Upton, New York 11973} 
  \author{H.~Atmacan}\affiliation{University of Cincinnati, Cincinnati, Ohio 45221} 
  \author{T.~Aushev}\affiliation{Higher School of Economics (HSE), Moscow 101000} 
  \author{R.~Ayad}\affiliation{Department of Physics, Faculty of Science, University of Tabuk, Tabuk 71451} 
  \author{V.~Babu}\affiliation{Deutsches Elektronen--Synchrotron, 22607 Hamburg} 
% ...
\collaboration{The Belle Collaboration}


\begin{abstract}

\par We report the results of the first search for the decay $\beep$ using $121.4~\textrm{fb}^{-1}$ of data 
collected at the $\Upsilon(5S)$ resonance 
with the Belle detector 
at the KEKB asymmetric-energy $e^+e^-$ collider. 
We observe no significant signal and set a 90\% confidence-level upper limit of 
$6.5 \times 10^{-5}$ 
on the branching fraction of this decay.

\end{abstract}




\maketitle

\tighten

{\renewcommand{\thefootnote}{\fnsymbol{footnote}}}
\setcounter{footnote}{0}


%

The charmless hadronic decay $\bseep$ is suppressed in the Standard Model (SM) 
and 
proceeds only through transitions 
sensitive to Beyond-the-Standard-Model (BSM) physics~\cite{Bevan:2014iga}. 
BSM scenarios, 
such  as  a  fourth  generation  of  fermions, 
supersymmetry with broken R-parity, 
and 
a two-Higgs doublet model with flavor-changing neutral currents, 
could affect the branching fraction and 
{\it CP} asymmetry of this decay~\cite{belleiiphysicsbook}.
The expected branching fraction 
for $\beep$ in the SM spans a range of
$(2 - 4)\times10^{-5}$~\cite{bf1, bf2, bf3, bf4, bf5}. 
Once branching fractions for two-body decays 
$B_{d,s}^0 \to \et\et$, $\etp\et$, and $\etp\etp $ are measured,
it would be possible to extract {\it CP}-violating parameters
using a formalism based on SU(3)/U(3) symmetry~\cite{bf1}.
To achieve this goal, at least four of these six branching fractions need to be measured.
Only the branching fraction for $B_s^0 \to \eta^{\prime}\eta^{\prime}$ has been measured so far ~\cite{bsepep}.

In this Letter, we report the results of the first search for the decay $\bseep$
using the full Belle data sample of $121.4~\textrm{fb}^{-1}$
collected at the $\Upsilon(5S)$ resonance.
The inclusion of the charge-conjugate decay mode is implied throughout. 
The Belle detector was
a large-solid-angle magnetic spectrometer
that operated at the KEKB asymmetric-energy $e^+e^-$ collider~\cite{KEKB}.
The detector components relevant to our study include
a tracking system comprising a silicon vertex detector (SVD) and a central drift chamber (CDC),
a particle identification (PID) system
that consists of a barrel-like arrangement of time-of-flight scintillation counters (TOF)
and an array of aerogel threshold Cherenkov counters (ACC),
and a CsI(Tl) crystal-based electromagnetic calorimeter (ECL).
All these components were located inside a superconducting
solenoid coil that provided a 1.5~T magnetic field.
A detailed description of the Belle detector can be found elsewhere~\cite{Belle}.

The $\Upsilon(5S)$ resonance decays into $\bsbs$  pairs, 
where the relative fractions of the two former decays are 
$\fbssbss =(87.0\pm1.7)\%$ and $\fbsbss=(7.3\pm1.4)\%$~\cite{frac}, respectively. 
Signal $\bs$ mesons originate 
from the direct decays of $\Upsilon(5S)$ 
or from radiative decays of the excited vector state $\bst$. 
The $\Upsilon(5S)$ production cross section is $340 \pm 16$~pb~\cite{frac}. 
To present our nominal result for $\mathcal{B}(\bseep)$ 
we use the world average value 
for the fraction of $\bsbsbara$ in $b\bar{b}$ events 
$f_s = 0.201 \pm 0.031$~\cite{PDG}, 
the data sample is therefore estimated to contain 
$(16.60 \pm 2.68) \times 10^{6}$ $\bs$ mesons. 
We also report the results for $f_s \times \mathcal{B}(\bseep)$. 

To maximize discovery potential of the analysis
and
to validate the signal extraction procedure, 
we use  a sample of background
Monte Carlo (MC) simulated events 
equivalent to six times the data statistics. 
In addition, 
to estimate the overall reconstruction efficiency 
we use a high-statistics signal MC sample, 
where the other $B_s^{(\ast)0}$ meson decays according to known branching fractions~\cite{PDG}. 
Both samples are used to develop
a model implemented
in the unbinned extended maximum-likelihood (ML) fit to data.
The MC-based model is validated with 
a control sample of $\bksep$ decays 
reconstructed from 711~${\rm fb^{-1}}$ 
of $\Upsilon(4S)$ data.

We reconstruct $\et$ candidates
using pairs of electromagnetic showers
not matched to the projections
of charged tracks to the ECL 
and therefore identified as photons.
We require that the reconstructed energies of these showers
exceed 50 (100) MeV in the barrel (endcap)
region of the ECL.   
The larger 
energy threshold for the endcaps 
is due to the larger beam-related background in these regions.
To reject hadronic showers mimicking photons,
the ratio of the energies 
deposited by a photon candidate in the $(3\times3)$ and $(5\times 5)$ 
ECL crystal arrays 
centered on the crystal with the largest deposited energy 
is required to exceed 0.75. 
The reconstructed invariant mass of the $\et$ candidates 
is required to be $515 \le M(\gamma\gamma) \le 580$~${\rm MeV}/{\it c}^2$,
which corresponds, approximately, to a $\pm3\sigma$  
Gaussian resolution window 
around the nominal $\et$ mass~\cite{PDG}.
To suppress 
combinatorial background arising due to low-energy photons, 
the magnitude of the cosine of the helicity angle 
($\cos\theta_{\textrm{hel}}$) 
is required to be less than 0.97, 
where 
$\theta_{\textrm{hel}}$ is the angle 
in the $\eta$ candidate's rest frame 
between the directions 
of its Lorentz boost from the laboratory frame 
and one of the photons. 

The $\etp$ candidates are 
formed by combining 
pairs of oppositely charged  pions 
with 
the $\et$ candidates.
We require the reconstructed $\etp$ invariant mass to be in the range
$920\le \mep \le 980$~${\rm MeV}/{\it c}^{2}$, which corresponds,
approximately, to the range $[-10,+6]\sigma$
of the Gaussian 
resolution, 
after performing a kinematic fit
constraining the reconstructed 
mass
of the 
$\et$ candidate
to the nominal $\et$ mass~\cite{PDG}.
To identify charged pion candidates,
the ratios of PID likelihoods,
$R_{i/\pi}={{\mathcal L}}_{i}/({\mathcal{L}}_{\pi}+{\mathcal{L}}_{i})$,
are used, where $L_{\pi}$  is the likelihood for the track 
being a pion, 
while $L_i$ is the corresponding likelihood 
for the 
kaon ($i=K$) or electron ($i=e$) hypotheses.
We require $R_{K/\pi}\le0.6$ and $R_{e/\pi}\le0.95$ for pion candidates.
The likelihood for each particle species is obtained by combining  information 
from CDC, TOF and ACC~\cite{nakano_pid}, and (for electrons only) ECL~\cite{eid}. 
According to MC studies,
these requirements reject
28\% of background, 
%
%
while the resulting efficiency loss is below 3\%. 
Charged pion tracks are required to originate from near the interaction point (IP) 
by restricting their distance of closest approach 
to the $z$ axis 
to be less than 
4.0~cm along the $z$ axis 
and 
0.3~cm perpendicular to it, 
respectively. 
The $z$ axis is opposite to the direction of the $e^+$ beam. 
These selection criteria suppress beam-related backgrounds and reject poorly reconstructed tracks.
To reduce systematic uncertainties associated with track reconstruction efficiency,
the transverse momenta of charged pions are required to be greater than 100~${\rm MeV}/{\it c}$. 

To identify $\bseep$ candidates we use 
(shown here in natural 
units) 
the beam-energy-constrained $\bs$ mass, 
%
%
$\mbc=\sqrt{E_{\rm beam}^2-p_{B_s}^2}$,
the energy difference, $\de=E_{B_s}-E_{\rm beam}$,
and the reconstructed invariant mass of the $\eta^\prime$,
where $E_{\rm beam}$, $p_{B_s}$ and $E_{B_s}$
are the beam energy,
the 
momentum and energy of the
$\bs$ candidate,
respectively.
All these quantities are 
calculated 
in the $e^+e^-$ center-of-mass frame.
To improve the $\de$ resolution, 
the $\etp$ candidates are further constrained to the nominal 
mass of $\etp$, 
though most of the improvement comes from the $\et$ mass constraint. 
Signal candidates are required to satisfy selection criteria $\mbc>5.3$~${\rm GeV}/{\it c}^2$ and $-0.4 \le \de \le0.3$~GeV.
In a Gaussian approximation, the $\de$ resolution is approximately 40~MeV.
Similarly, 
the $\mbc$ 
resolution is 4~${\rm MeV}/c^2$. 
To take advantage of all available information 
in case the data indicate signal presence, 
we include 
$\mep$ in the three-dimensional (3D) ML fit 
used to statistically separate the signal from background. 
We define the signal region: 
$5.35<\mbc<5.43$~${\rm GeV}/{\it c}^2$, $-0.25<\de<0.10$~GeV, 
and $0.94<\mep<0.97$~${\rm GeV}/{\it c}^2$. 
The area outside the signal region is considered as sideband.
To optimize sensitivity we use a narrower signal region 
$5.39<\mbc<5.43$~${\rm GeV}/{\it c}^2$ 
which would contain the largest signal contribution. 

Hadronic continuum events from $e^+e^-\to q\bar{q}$ ($q=u,d,c,s$) 
are the primary source of background.
Because of large initial momenta of the light quarks,
continuum events exhibit a ``jetlike'' event shape,
while $\bsbsbara$ events are distributed isotropically.
We utilize modified Fox-Wolfram moments~\cite{ksfw},
used to describe the event topology,
to discriminate between signal and continuum background.
A likelihood ratio ($\mathcal{LR}$) is calculated
using Fisher discriminant coefficients obtained
in an optimization based on these moments.
We suppress the background
using a discovery-optimized selection on $\mathcal{LR}$
obtained by maximizing 
the value of Punzi's figure of merit~\cite{punzi}:

\begin{equation}
{\rm FOM} =\frac{\varepsilon(t) }{a/2+\sqrt{B(t)}},
\label{eq:FOM}
\end{equation}

\noindent where
$t$ is the requirement on $\mathcal{LR}$,
$\varepsilon$ and $B$ are the signal reconstruction efficiency 
and the number of background events expected in the signal region 
for a given value of $t$, respectively. 
The quantity $a$ is the desired significance 
(which we vary between 3 and 5) 
in the units of standard deviation. 
To predict $B(t)$ we multiply 
the number of events in the data sideband by 
the ratio of the numbers of events 
in the signal region and sideband 
in the background MC sample. 
We require signal candidates to satisfy the requirement $\mathcal{LR} \ge 0.95$, 
which corresponds to $B(0.95)=3.3$ and 48 background events in the
signal region and sideband, respectively. 
This 47\%-efficient 
requirement 
removes 99\% of background. 
Using MC simulation 
we estimate 
continuum background to comprise 
97\% 
of the remaining events. 

The background events containing real $\etp$ mesons exhibit a peak
in the $\mep$ distribution,
however, they are distributed  
smoothly 
in $\mbc$ and $\de$.
The fraction of this peaking background
is a free parameter in our ML fits.

About 14\% of the reconstructed signal MC events contain multiple candidates 
primarily arising 
due to misreconstructed $\et$ mesons.
In such events we retain the candidate 
with the smallest value of $\sum{\chi^2_{\eta}}+\chi^2_{\pi^+\pi^-}$, 
where $\chi^2_{\eta}$ denotes the $\et$ mass-constrained fit statistic, 
the summation is over the two $\et$ candidates, 
and $\chi^2_{\pi^+\pi^-}$ quantifies the quality of the vertex fit for two pion tracks. 
Simulation shows that this procedure selects the correct $\bs$ candidate in 62\% of such events. 
The overall reconstruction efficiency is 10\%. 

To extract the signal yield, we perform an unbinned extended ML fit
to the 3D distribution of $\mbc$, $\de$, and $\mep$.
The likelihood function is

\begin{equation}
\mathcal{L}=\frac{e^{-\sum_{j}^3 n_j}}{N!}\prod_{i=1}^{N}\left(\sum_{j}^3 n_{j}\mathcal{P}_{j}[M_{\rm bc}^i, \Delta E^i, M^i(\pi^+\pi^-\eta) ]\right), 
\label{eq:PDF}                                                                                                         
\end{equation} 

\noindent where $i$ is the event index, 
$N$ is the total number of events, 
$j$ denotes the fit component 
(the three components are 
background, 
correctly reconstructed signal, 
and 
misreconstructed signal 
described later), 
and the parameters $n_j$ represent 
signal and background yields. 
Due to negligible correlations among fit variables for both
background and correctly reconstructed signal events,  
the probability density function (PDF) 
for each fit component 
is assumed to factorize as 
$\mathcal{P}[M_{\rm bc}^i, \Delta E^i, M^i(\pi^+\pi^-\eta) ] = 
\mathcal{P}[\mbc^{i}] \cdot \mathcal{P}[\de^{i}] \cdot \mathcal{P}[M^i(\pi^+\pi^-\eta)]$. 
The signal PDF 
is represented by a weighted sum of the three PDFs 
describing possible $\bseep$ signal contributions from $\bsbsbara$ pairs,
where the weights are fixed according to previous measurements~\cite{frac}.

To validate our fitting model 
and 
adjust the PDF shape parameters used to describe the signal,
we use the control sample of $\bksep$ decays. 
We reconstruct $K_S^0$ candidates via secondary vertices 
associated with pairs of oppositely charged pions~\cite{ks_reco}
using a neural network technique~\cite{NN}. 
The following information is used in the network: 
the momentum of the $K_S^0$ candidate in the laboratory frame;
the distance along the $z$ axis between
the two track helices at the point of their closest approach;
the flight length in the $x-y$ plane;
the angle between the $K_S^0$ momentum
and the vector joining the $K_S^0$ decay vertex to the IP;
the angle between the pion momentum  and the laboratory-frame $K_S^0$ momentum in the $K_S^0$ rest frame;
the distance-of-closest-approach in the $x-y$ plane between the IP and the two pion helices;
and the pion hit information in the SVD and CDC.
The selection efficiency is 87\% over the momentum range of interest.
We also require that the reconstructed $\pi^+\pi^-$ invariant mass
is within 12~${\rm MeV}/{\it c}^2$,  
which is about 3.5$\sigma$, 
of the nominal $K_S^0$ mass~\cite{PDG}.
We require $5.20\le \mbc \le 5.30$~${\rm GeV}/{\it c}^2$ for $B^0$ candidates.
The control-sample signal region is 
$5.27<\mbc<5.29$~${\rm GeV}/{\it c}^2$, $-0.20<\de<0.10$~GeV, 
and $0.94<\mep<0.97$~${\rm GeV}/{\it c}^2$. 
All other selection criteria  
are the same as those used to select $B_s^0$ candidates.
This control sample is used to validate 
the $\eta$ and $\eta^\prime$ reconstruction 
and its effect on the resolution functions and 
PDF shape parameters. The validation of $K_S^0$ reconstruction 
was performed previously in a similar $B_s^0$ analysis~\cite{Pal:2015ghq}.

The presence of four photons in the final state gives rise to
a sizable misreconstruction probability for the signal events.
We study these self-crossfeed (SCF) events using the signal MC sample. 
A large correlation between $\mbc$ and $\de$ for such signal events
is taken into account by describing the correctly reconstructed signal
and SCF components separately with two different PDF sets. 
The latter comprise approximately 14\% of the reconstructed signal 
and are excluded from the estimate of its efficiency. 
The Pearson correlation coefficient 
for the region with 
largest 
correlations 
for SCF signal events is 27\%.

A sum of a Gaussian and a Crystal Ball~\cite{xbal} function 
is used to model the correctly reconstructed signal in each of the three fit variables. 
For $\mbc$ and $\mep$ we use a sum of these two functions with the same mean 
but different widths, while for $\de$ both the mean and width are different. 
A Bukin function~\cite{bukin} and an asymmetric Gaussian are used to model the SCF contribution
in $\mbc$ and $\de$, respectively.
For $\mep$, we use a sum of a Gaussian and
a first-order Chebyshev polynomial.
In our nominal fit to data the fraction of correctly reconstructed signal is fixed to its MC value.  
The signal PDF shape parameters for $\mbc$ and $\de$ 
are validated using the $\bksep$ control sample.

\begin{figure*}[htb!]
\small
\begin{center}
\includegraphics[width=\textwidth]{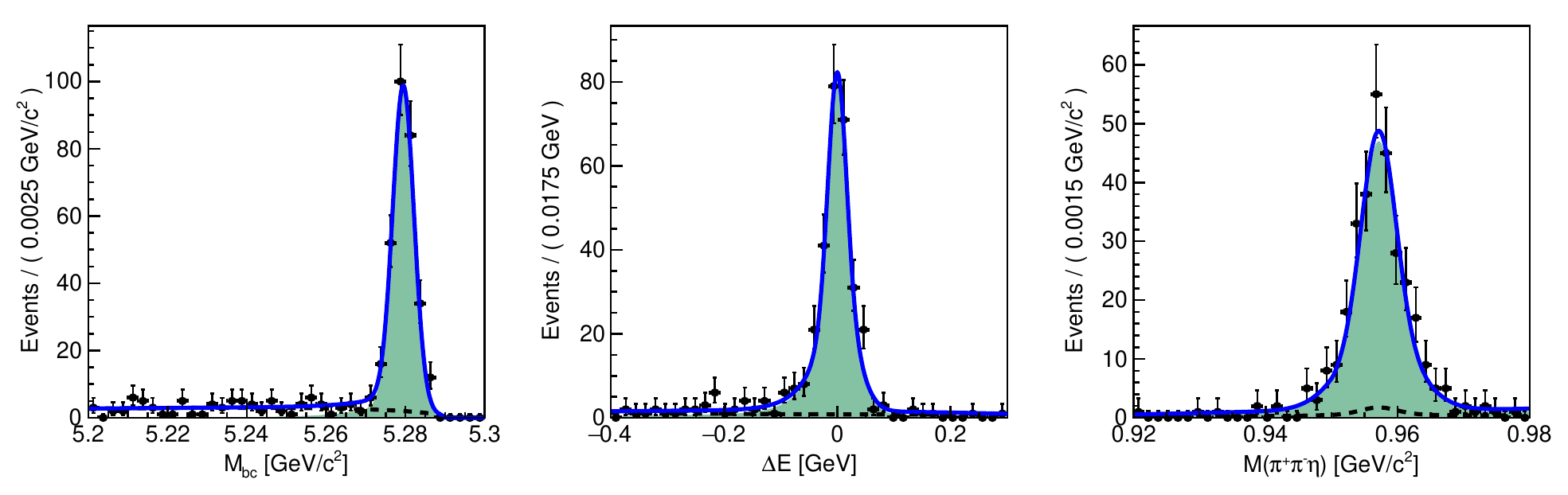}
\end{center}
\caption{Signal-region projections of the fit results on $\mbc$, $\de$, and $\mep$ for the $\bksep$ control sample. 
Points with error bars are data, 
blue solid curves are the results of the fit, 
black dashed curves are the background component, and 
cyan-filled regions show the signal component. 
}
\label{fit_data_y4s}
\end{figure*}

We use an ARGUS~\cite{argus} function to describe the background distribution in $\mbc$ 
and 
a first-order Chebyshev polynomial for $\de$. 
To model the peaking part in $\mep$ we use the signal PDF, 
because the peak is due to real $\etp$ mesons,
while an additional first-order Chebyshev polynomial is used for the non-peaking contribution.
The projections of the fit to the $\bksep$ control sample 
are shown in Fig.~\ref{fit_data_y4s}.

\begin{figure*}[htb!]
\small
\begin{center}
\includegraphics[width=\textwidth]{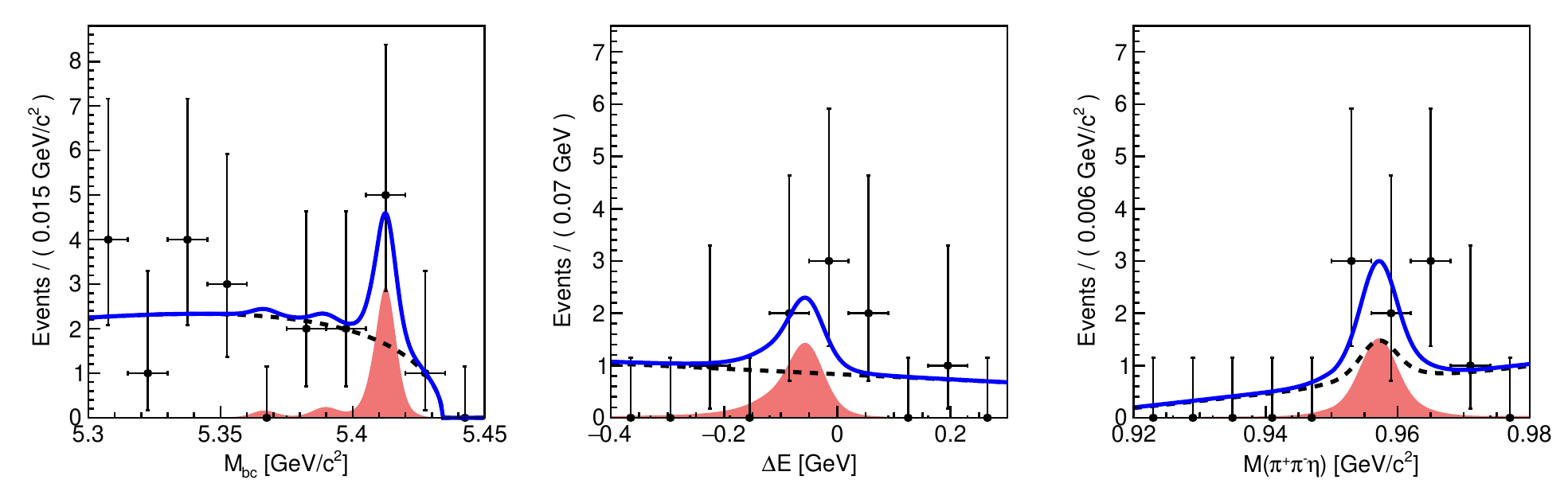}
\end{center}
\caption{Signal-region projections of the fit results on $\mbc$, $\de$, and $\mep$  for $\bseep$. 
The $\mbc$ signal region of the dominant signal contribution, 
$5.39 < \mbc < 5.43$~${\rm GeV}/{\it c}^2$, 
is used to plot the $\de$ and $\mep$ projections. 
Points with error bars are data, 
blue solid curves are the results of the fit, 
black dashed curves are the background component, and 
pink-filled regions show the signal component. 
The three $\mbc$ peaks in the signal component 
(from right to left) 
correspond to contributions from 
$\bsbs$ 
pairs. 
}
\label{fit_data}
\end{figure*}

To further test and validate our fitting model,
ensemble tests are carried out by generating MC pseudoexperiments.
In these experiments we use PDFs obtained from full detector simulation and the $\bksep$ data.
We perform 
1000 pseudoexperiments for each assumed number of signal events.
An ML fit is executed for each sample prepared in these experiments.
The signal yield distribution obtained from these fits 
exhibits good linearity. 
We use the results of pseudoexperiments
to construct classical confidence intervals (without ordering)
using a procedure due to Neyman~\cite{frequentist_approach}.
For each ensemble of pseudoexperiments,
the lower and upper ends of the respective confidence interval
represent the values of fit signal yields
for which 10\% of the results lie below and above these values,
respectively.
These intervals are then combined to prepare
a classical confidence belt~\cite{belt_method,belt_method_2} 
used to make a statistical interpretation
of the results obtained from data. 
The confidence intervals prepared using this statistical method 
are known to slightly ``overcover'' 
for the number of signal events~\cite{fc}, 
therefore resulting in a conservative upper limit. 

We apply the 3D model to the data and
obtain $2.7 \pm 2.5$ signal and $57.3 \pm 7.8$ background events.
The signal-region projections of the fit are shown in Fig.~\ref{fit_data}.
We observe no significant signal and estimate a 90\% confidence-level (CL) upper limit on the branching fraction
for the decay $\bseep$ 
using 
the following formula:

\begin{equation}
\mathcal{B}(\bseep) < \frac{N_{\textrm{UL}}^{90\%}}{N_{\bs} \times \varepsilon \times \mathcal{B}}~,
\label{eq_ul}
\end{equation}

\noindent where $N_{\bs}$ is the number of $\bs$ mesons 
in the full Belle data sample,
$\varepsilon$ is the overall reconstruction efficiency for the signal $B_s^0$ decay, 
and $\mathcal{B}$ 
is the product of the 
subdecay 
branching fractions 
for $\eta$ and $\eta^\prime$ 
reconstructed in our analysis. 
Further, $N_{\textrm{UL}}^{90\%}$ is the expected signal 
yield of approximately 6.6 events at 90\% CL obtained from the confidence belt 
constructed using the frequentist approach~\cite{frequentist_approach}. 
Using Eq.~(\ref{eq_ul}) 
we estimate a 90\% CL upper limit 
on the branching fraction 
$\mathcal{B}(\bseep) < 6.2 \times 10^{-5}$. 
We also estimate a 90\% CL upper limit 
on the product 
$f_s \times \mathcal{B}(\bseep) < 1.2 \times 10^{-5}$. 
The systematic uncertainties are not included in these estimates. 

Sources of systematic uncertainties and their relative contributions 
are listed in Table~\ref{tab:lkr_sys}. 
The relative uncertainties 
on 
$f_s$ 
and 
$\sigma(\Upsilon(5S))$ 
are 
15.4\% 
and 
4.7\%, 
respectively. 
The systematic uncertainty due to $\eta$ reconstruction is 2.1\% per $\et$ candidate~\cite{eta_syst}. 
Track reconstruction~\cite{track_syst} and PID 
systematic uncertainties are 0.35\% and 2\% per track, respectively. 
We estimate the systematic uncertainty 
due to the $\mathcal{LR}$ requirement to be 10\%, 
which represents the relative change in efficiency 
when this requirement is varied by $\pm$0.02 
about the nominal value of 0.95. 
This range of variation is defined by the statistics of the control sample 
which is used to validate the efficiency and 
its dependence on the $\mathcal{LR}$ requirement. 
Systematic uncertainty due to signal PDF shape is estimated 
by varying the fixed parameters 
within their statistical uncertainties 
determined with $\bksep$ data. 
When varying these parameters, we observe 
an 11\% change in the signal yield obtained from the data 
and use this number as an estimate of PDF parametrization systematics. 
Systematic uncertainty due to $\fbssbssall$ is evaluated 
by varying relative fractions of possible contributions to signal PDF and is 1.3\%. 
When varying the SCF contribution by $\pm 50$\% of itself, 
we observe a 4\% change in the results of the fit to data, 
which we use as an estimate of SCF PDF systematic uncertainty. 
The relative uncertainties on 
$\et$ and $\etp$ branching fractions 
are 1\% and 1.2\%, respectively.
The statistical uncertainty due to MC statistics 
is estimated to be 0.1\%. 
The overall systematic uncertainties for 
$\mathcal{B}(\bseep)$ 
and 
$f_s \times \mathcal{B}(\bseep)$ 
are estimated by adding the individual contributions in quadrature and are 
23.1\% and 17.2\%, 
respectively. 
These systematic uncertainties are included in the 
$N_{\textrm{UL}}^{90\%}$ estimates 
of approximately 7.0 and 6.9 events 
by smearing the fit yield distributions 
while constructing the confidence belt 
used to extract the results. 
We 
estimate 
the upper limits 
on the branching fraction 
$\mathcal{B}(\bseep) < 6.5 \times 10^{-5}$ 
and 
on the product 
$f_s \times \mathcal{B}(\bseep) < 1.3 \times 10^{-5}$ 
at 90\% CL. 
Finally, using the number of signal events obtained from the fit 
we estimate 
$\mathcal{B}(\bseep) = (2.5 \pm 2.2 \pm 0.6) \times 10^{-5}$ 
and 
$f_s \times \mathcal{B}(\bseep) = (0.51 \pm 0.44 \pm 0.09) \times 10^{-5}$, 
where, for each of the two estimates, 
the first uncertainty is statistical and the second is systematic. 
We summarize the results in Table~\ref{tab:results}. 

\begin{table}
\caption{Summary of systematic uncertainties.}
\begin{ruledtabular}
\begin{tabular}{l|c}
Source     & Uncertainty (\%) \\
\hline
$f_s$                   & 15.4 \\ 
$\sigma(\Upsilon(5S))$  & 4.7  \\
$\eta$  reconstruction                 & 4.2  \\
Tracking                               & 0.7  \\
PID                                    & 4.0  \\
$\mathcal{LR}$ selection               & 10.0 \\
PDF parametrization                    & 11.0 \\
$\fbssbssall$                          & 1.3  \\
SCF PDF                                & 4.0  \\
Branching fraction of $\et$            & 1.0  \\
Branching fraction of $\etp$           & 1.2  \\
MC statistics                          & 0.1  \\
\end{tabular}
\end{ruledtabular}
\label{tab:lkr_sys}
\end{table}

\begin{table}
\caption{Summary of the results for 
$f_s \times \mathcal{B}(\bseep)$ and 
$\mathcal{B}(\bseep)$.  
See the text for more information.}
\begin{ruledtabular}
\begin{tabular}{c|c}
Quantity     & Value \\
\hline
\multirow{2}{*}{$f_s \times \mathcal{B}(\bseep)$} &  $(0.51 \pm 0.44 \pm 0.09) \times 10^{-5}$ \\
                                 & $< 1.3 \times 10^{-5}$ @ 90\% CL \\
\hline
\multirow{2}{*}{$\mathcal{B}(\bseep)$}   &  $(2.5 \pm 2.2 \pm 0.6) \times 10^{-5}$ \\
                        & $< 6.5 \times 10^{-5}$ @ 90\% CL \\
\end{tabular}
\end{ruledtabular}
\label{tab:results}
\end{table}

In summary, we have used the full data sample recorded
by the Belle experiment at the $\Upsilon(5S)$ resonance 
to search for the decay $\bseep$. 
We observe no statistically significant signal and set a 90\% CL upper 
limit of $6.5 \times 10^{-5}$ on its branching fraction.
To date, our result is the only experimental information on $\bseep$ 
and is twice as large as the most optimistic SM-based
theoretical prediction. 
This decay can be probed further at the 
next-generation Belle~II experiment~\cite{belle2} 
at the SuperKEKB collider in Japan.

We thank the KEKB group for the excellent operation of the
accelerator; the KEK cryogenics group for the efficient
operation of the solenoid; and the KEK computer group, and the Pacific Northwest National
Laboratory (PNNL) Environmental Molecular Sciences Laboratory (EMSL)
computing group for strong computing support; and the National
Institute of Informatics, and Science Information NETwork 5 (SINET5) for
valuable network support.  We acknowledge support from
the Ministry of Education, Culture, Sports, Science, and
Technology (MEXT) of Japan, the Japan Society for the 
Promotion of Science (JSPS), and the Tau-Lepton Physics 
Research Center of Nagoya University; 
the Australian Research Council including grants
DP180102629, 
DP170102389, 
DP170102204, 
DP150103061, 
FT130100303; 
Austrian Federal Ministry of Education, Science and Research (FWF) and
FWF Austrian Science Fund No.~P~31361-N36;
the National Natural Science Foundation of China under Contracts
No.~11435013,  
No.~11475187,  
No.~11521505,  
No.~11575017,  
No.~11675166,  
No.~11705209;  
Key Research Program of Frontier Sciences, Chinese Academy of Sciences (CAS), Grant No.~QYZDJ-SSW-SLH011; 
the  CAS Center for Excellence in Particle Physics (CCEPP); 
the Shanghai Pujiang Program under Grant No.~18PJ1401000;  
the Shanghai Science and Technology Committee (STCSM) under Grant No.~19ZR1403000; 
the Ministry of Education, Youth and Sports of the Czech
Republic under Contract No.~LTT17020;
Horizon 2020 ERC Advanced Grant No.~884719 and ERC Starting Grant No.~947006 ``InterLeptons'' (European Union);
the Carl Zeiss Foundation, the Deutsche Forschungsgemeinschaft, the
Excellence Cluster Universe, and the VolkswagenStiftung;
the Department of Atomic Energy (Project Identification No. RTI 4002) and the Department of Science and Technology of India; 
the Istituto Nazionale di Fisica Nucleare of Italy; 
National Research Foundation (NRF) of Korea Grant
Nos.~2016R1\-D1A1B\-01010135, 2016R1\-D1A1B\-02012900, 2018R1\-A2B\-3003643,
2018R1\-A6A1A\-06024970, 2018R1\-D1A1B\-07047294, 2019K1\-A3A7A\-09033840,
2019R1\-I1A3A\-01058933;
Radiation Science Research Institute, Foreign Large-size Research Facility Application Supporting project, the Global Science Experimental Data Hub Center of the Korea Institute of Science and Technology Information and KREONET/GLORIAD;
the Polish Ministry of Science and Higher Education and 
the National Science Center;
the Ministry of Science and Higher Education of the Russian Federation, Agreement 14.W03.31.0026, 
and the HSE University Basic Research Program, Moscow; 
University of Tabuk research grants
S-1440-0321, S-0256-1438, and S-0280-1439 (Saudi Arabia);
the Slovenian Research Agency Grant Nos. J1-9124 and P1-0135;
Ikerbasque, Basque Foundation for Science, Spain;
the Swiss National Science Foundation; 
the Ministry of Education and the Ministry of Science and Technology of Taiwan;
and the United States Department of Energy and the National Science Foundation.


\end{document}